\documentclass[12pt]{article}

\input{epsf}
\usepackage{subfigure}
\usepackage{cite}
\usepackage[dvips]{graphicx}
\usepackage{amsmath}
\usepackage{epsfig}
\usepackage{amsfonts}
\usepackage{amssymb}
\usepackage{multirow}
\usepackage{array}

\usepackage{pstricks}
\usepackage{color}

\usepackage{epsfig}
\usepackage{amsmath}
\usepackage{amssymb}

\usepackage{subfigure}

\newcommand{\be}{\begin{equation}}
\newcommand{\ee}{\end{equation}}
\newcommand{\bee}{\begin{equation*}}
\newcommand{\eee}{\end{equation*}}
\newcommand{\bea}{\begin{eqnarray}}
\newcommand{\eea}{\end{eqnarray}}
\newcommand{\bean}{\begin{eqnarray*}}
\newcommand{\eean}{\end{eqnarray*}}

\newcommand{\nn}{\nonumber}

\newcommand{\lp}{\left(}
\newcommand{\rp}{\right)}

\usepackage{amsfonts}
\usepackage{amssymb}
\usepackage{amsmath}
\usepackage{mathrsfs}
\usepackage{latexsym}
\usepackage{cancel}
\usepackage{bm}
\usepackage{bbm}
\usepackage[mathcal]{euscript}
\usepackage{graphicx}
\usepackage{color}

\begin{document}

\setcounter{page}{0}
\thispagestyle{empty}

\begin{flushright}
CERN-PH-TH/2011-086\\
\today
\end{flushright}

\vskip 8pt

\begin{center}
{\bf \LARGE {
Cosmological Consequences of 
\vskip 8pt
Nearly Conformal Dynamics at the TeV scale
 }}
\end{center}

\vskip 12pt

\begin{center}
 {\bf Thomas Konstandin$^{a}$ and G\'eraldine  Servant$^{a,b}$ }
\end{center}

\vskip 20pt

\begin{center}

\centerline{$^{a}${\it CERN Physics Department, Theory Division, CH-1211 
Geneva 23, Switzerland}}
\centerline{$^{b}${\it Institut de Physique Th\'eorique, CEA/Saclay, F-91191 
Gif-sur-Yvette C\'edex, France}}
\vskip .3cm
\centerline{\tt tkonstan@cern.ch, geraldine.servant@cern.ch}
\end{center}

\vskip 13pt

\begin{abstract}
\vskip 3pt
\noindent

Nearly conformal dynamics at the TeV scale as motivated by the hierarchy
problem can be characterized by a stage of significant supercooling at the
electroweak epoch.  This has important cosmological consequences.  In
particular, a common assumption about the history of the universe is
that the reheating temperature is high, at least high enough to assume
that TeV-mass particles were once in thermal equilibrium. However, as
we discuss in this paper, this assumption is not well justified in some
models of strong dynamics at the TeV scale. We then need to reexamine
how to achieve baryogenesis in these theories as well as reconsider
how the dark matter abundance is inherited.  We argue that baryonic and
dark matter abundances can be explained naturally in these setups
where reheating takes place by bubble collisions at the end of the
strongly first-order phase transition characterizing conformal
symmetry breaking, even if the reheating temperature is below the
electroweak scale $\sim 100$ GeV. We also discuss inflation as well as
gravity wave smoking gun signatures of this class of models.

\end{abstract}

\newpage

\tableofcontents

\vskip 13pt

\section{Introduction\label{sec:intro}}

Within the next few years, the LHC will be probing the electroweak (EW)
symmetry breaking sector of the Standard Model (SM). In the SM, there
is no understanding of the dynamics responsible for the breaking of
the $SU(2)_L\times U(1)_Y$ gauge invariance.  Electroweak symmetry
breaking is
introduced by hand through the addition of the Higgs mass operator.

There are two main avenues for explaining the lightness of the scalar
Higgs boson: supersymmetry and Higgs compositeness.  Minimal
supersymmetry predicts a too light Higgs unless a tuning is invoked.
The idea of Higgs compositeness has therefore received a revival of
interest in the last few years~\cite{Agashe:2004rs,Contino:2010rs}.
In this framework, EW symmetry breaking is triggered by a
spontaneously broken nearly scale invariant sector at a scale
$\Lambda_{\rm CFT} \sim 4 \pi f \geq \Lambda_{\rm EW}\sim 4 \pi v$.
The hierarchy $\Lambda_{\rm EW} \ll M_{\rm Pl}$ is explained
dynamically via dimensional transmutation, as the quantum running of a
dimensionless coupling generates a new scale, like in QCD.  The
presence of a moderate separation between the scale of EW symmetry
breaking $v=246$ GeV and the scale of conformal breaking $f$ allows to
keep under control the unwanted corrections to EW precision
observables and is conceivable if the dynamics responsible for EW
symmetry breaking is strongly coupled and nearly conformal, like in
theories of walking technicolor~\cite{Holdom:1981rm} or via AdS/CFT in
Randall-Sundrum extra-dimensional warped
geometries~\cite{Randall:1999ee,ArkaniHamed:2000ds,Rattazzi:2000hs}.
The spectrum of states at the EW scale in these theories contains the
pseudo-Goldstone dilaton from the spontaneous breaking of conformal
invariance (the radion in the 5D picture).

Extensive studies have been devoted to the phenomenology of this
scenario, in particular to make it consistent with experimental
constraints and determine its distinctive signatures at colliders, see
e.g.~Ref.~\cite{Contino:2006nn}.  Somehow, the cosmological
consequences of this framework have not been much explored although
they can be completely different from the standard picture
~\cite{Creminelli:2001th, Randall:2006py, Nardini:2007me,
  Konstandin:2010cd, Kaplan:2006yi} and open new and unique avenues
for addressing the main cosmological puzzles such as the nature of
dark matter, the origin of the matter-antimatter asymmetry and
inflation\footnote{Literature exists on the possibility that dark matter is made of composite states or on baryogenesis proposals based on technibaryons, however it generally assumes a standard cosmological evolution, e.g.~\cite{Nussinov:1985xr,Chivukula:1989qb}.}. 
The goal of this paper is to present some general
properties which we believe are quite typical in a large class of
theoretically well-motivated models.  We will argue that in a certain
class of theories of EW symmetry breaking with nearly conformal
dynamics, we have the following:
\begin{enumerate}
\item A period of supercooling with some efolds of inflation due to a
  strongly first-order phase transition is likely.
\item The reheat temperature of the universe is then at the EW scale
  and depending on the mass of the radion and Higgs, it can even be
  below the typical sphaleron freeze-out temperature ($T_{\rm reh}
  \lesssim T_{\rm sph}\sim 120$ GeV.)

  Therefore, dilution of particle abundances during the short
  inflationary stage and a low reheat temperature both require to
  re-examine dark matter production and baryogenesis.

\item A large signal in gravity waves in the millihertz range is a
  smoking-gun signature of this scenario and therefore of
  nearly conformal dynamics at the TeV scale. For any other model with a
  strongly first-order phase transition at the EW scale, a detectable
  stochastic background of gravitational waves is fine-tuned.

\end{enumerate}

The main distinctive feature of the cosmological scenario we are
considering is two-fold: Reheating comes from bubble collisions
following conformal symmetry breaking after the universe has undergone
a stage of significant supercooling.  Along, the reheat temperature of
the universe is close to the scale of the induced EW symmetry
breaking, and depending on its precise value, it will restrict the
range of viable mechanisms for both visible and dark matter genesis.
Our underlying framework is therefore very different from the common
cosmological paradigm where the reheat temperature of the universe is
assumed to be well above the EW scale, which on the other hand is
rather natural if inflation occurs at a scale $\sim 10^{15}$ GeV.
However, there is no real need for a reheat temperature much above the
EW scale.  As we will argue, it is possible to explain the visible and
dark matter abundances even with a reheat temperature below the weak
scale.

The paper is organized as follows. In Section \ref{sec:PT}, we present
the main properties on which we rely and study the nature of the phase
transition. We also estimate the number of efolds, summarize all the
constraints and consider inflation. Section \ref{sec:reh} is about
reheating and consequences for baryogenesis and dark matter particle
production. We discuss experimental probes in Section \ref{sec:exp},
and we conclude in Section~\ref{sec:dis}.

\section{A generic strongly first-order phase transition at 
the electroweak scale \label{sec:PT}}

In most of this paper, we will not make any particular assumptions
about the nature of the strongly interacting sector.  Our whole
discussion will rely on the general assumption that the scalar
effective potential describing symmetry breaking is a scale invariant
function modulated by a slow evolution:
\begin{equation}
V(\mu)=\mu^4 P\left[ \ \left(\frac{\mu}{\mu_0}\right)^{\epsilon} \
\right], \label{equation:nearlyconformal} 
\end{equation} 
similarly to the Coleman-Weinberg potential where a slow RG evolution
of the potential parameters can generate very separated scales.  $P$
is a polynomial function reflecting some explicit breaking of
conformal invariance by turning on some coupling of dimension
$-\epsilon$. This potential generically has a minimum at $\mu_- \neq 0
$. We are interested in the case where $|\epsilon |$ is small so that
we have an almost marginal deformation of the CFT. If $\epsilon>0$
symmetry breaking results from a balance between two operators unlike
in QCD where it is driven by the blow-up of the gauge
coupling~\cite{ArkaniHamed:2000ds, Rattazzi:2000hs}. For $|\epsilon|
\ll 1$, a large hierarchy is generated.

\subsection{Cosmological properties of a nearly conformal scalar potential}

\begin{figure}[t]
\begin{center}
\includegraphics[width=0.575\textwidth, clip ]{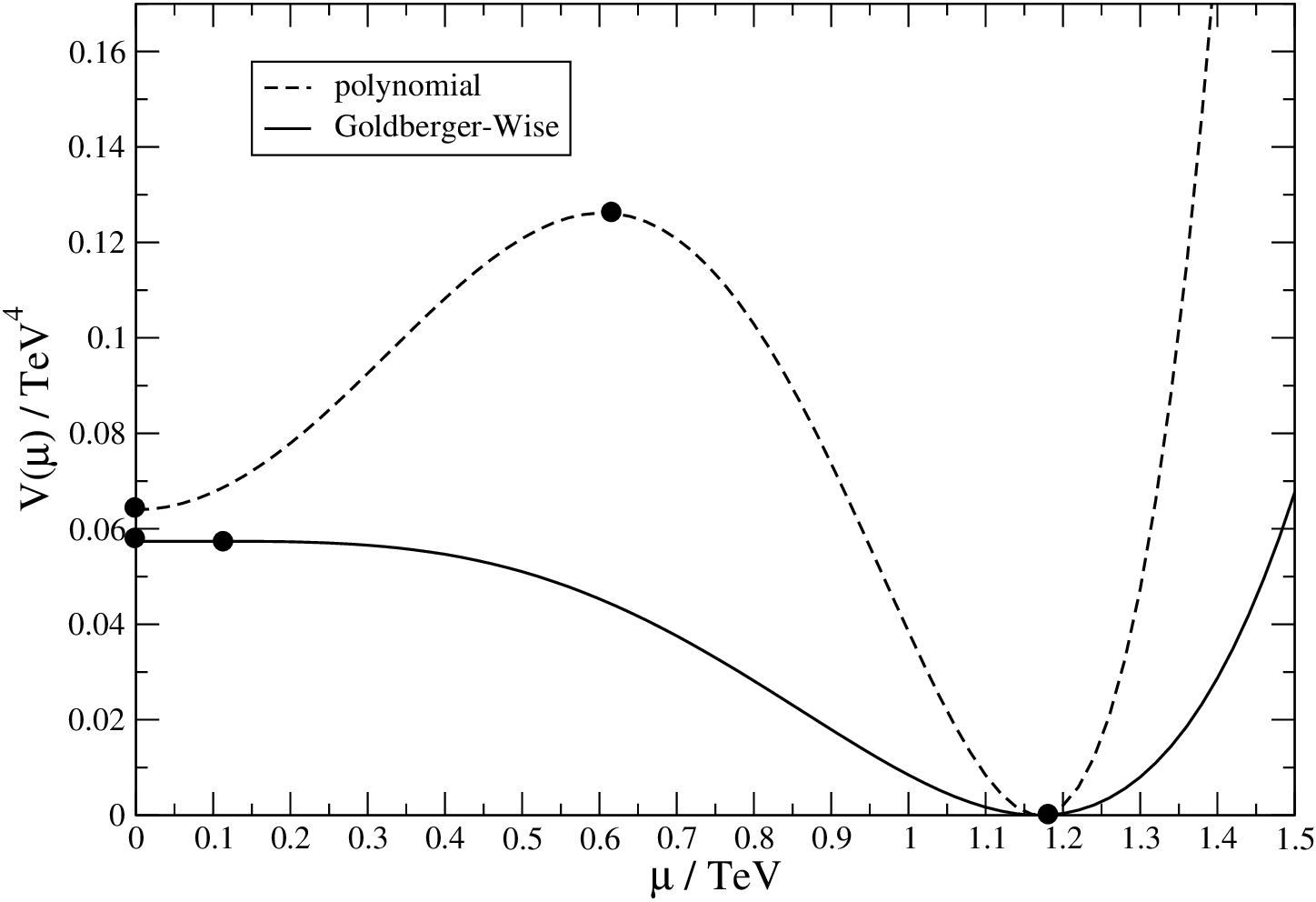}
\caption{\label{fig:potentials} \small Comparison of a typical
polynomial potential given here by $\lambda(\mu^2
-\mu_0^2)^2 +\frac{1}{\Lambda^2} (\mu^2
-\mu_0^2)^3$ with a nearly conformal potential of the type
of eq.~(\ref{equation:nearlyconformal}).  Both have a minimum at
$\mu_{\rm min}\sim 1.2$ TeV.  For the usual polynomial potential
$\mu_{\rm max}/ \mu_{\rm min} \sim {\cal O}(1)$, unless coefficients are
fine-tuned while for the potential (\ref{equation:nearlyconformal})
with $|\epsilon| < 1$, one can easily get a shallow potential with
widely separated extrema. In this particular example
$|\epsilon|=0.2$. The $\bullet$ indicates the position of the
maxima.}
\end{center}
\end{figure}

This class of potentials leads to some unique cosmological properties.
In particular, it leads to a strongly first-order phase transition.
What makes the nearly conformal potentials special is the fact that the
positions of the maximum $\mu_+$ and of the minimum $\mu_-$ can be
very far apart in contrast with standard polynomial potentials where
they are of the same order, as illustrated in
Fig.~\ref{fig:potentials}. This makes the temperature dependence of
the tunneling action behave very differently from the case of standard
polynomial potentials.  The nucleation temperature $T_n$ is determined
by the tunneling point $\mu_r$ (also called {\it release point}),
which is located behind the barrier, somewhere between the maximum and
the minimum of the potential.  For a standard polynomial potential,
$\mu_+$ and $\mu_-$ are of the same order and the tunneling point is
of the same order as the value of the field at the minimum of the
potential. For a nearly conformal potential, the two extrema are
widely separated and as we will show, the release point can be as low
as $\mu_r \gtrsim \sqrt{\mu_+ \mu_-} \ll \mu_-$. Since the nucleation
temperature $T_n \propto \mu_r$, we can get a very small $T_n$
compared to the vacuum expectation value of the scalar field $\mu_-$
and therefore several efolds of inflation.

Typically, an extended phase of inflation (at least several efolds)
cannot be ended by a first-order phase transition. This is the
well-known graceful exit problem of old inflation 
which results from the following argument: for a generic free energy
$V(\phi, T)$ the tunnel action $S_3/T$ is a ``well-behaved" (meaning
roughly polynomial) function of the temperature $T$. The first
nucleated bubbles appear when the temperature satisfies, in terms of
the Hubble constant $H$,
\be
S_3 / T \approx \log \frac{T^4}{H^4}.
\ee
At the weak scale, this corresponds to $S_3 / T \approx 140$.  In
order to realize several efolds of inflation, the onset of the phase
transition and bubble nucleation should happen at a temperature $T_n$
that is several orders of magnitude smaller than the critical
temperature $T_c$ defined as the temperature at which the symmetric
and broken phase are degenerate.

If $S_3$ is a well-behaved function of $T$, characterized by the
energy scale $\mu_0\sim T_c$, its derivative $\partial_T (S_3/T)$ is
likewise and the parameter $\beta$ which quantifies the inverse
duration of the phase transition satisfies
\be
\beta / H =\left.  T \frac{d}{dT} \frac{S_3}{T} \right|_{T_n}
 \sim \frac{T_n}{\mu_0} \ \left.  \frac{S_3}{T} \right|_{T_n}. 
\ee
An extended phase of inflation (for example, $N_{\rm efolds}\sim
\log{T_c/T_n}\sim 10$ $\rightarrow$ $T_n/T_c \sim 10^{-4}$)
corresponds to $T_n \ll \mu_0$ then $\beta/H\ll 1 $, which implies
that bubbles never percolate and the phase transition cannot complete
and reheating never occurs.

\begin{figure}[t!]
\begin{center}
\includegraphics[width=0.625\textwidth, clip ]{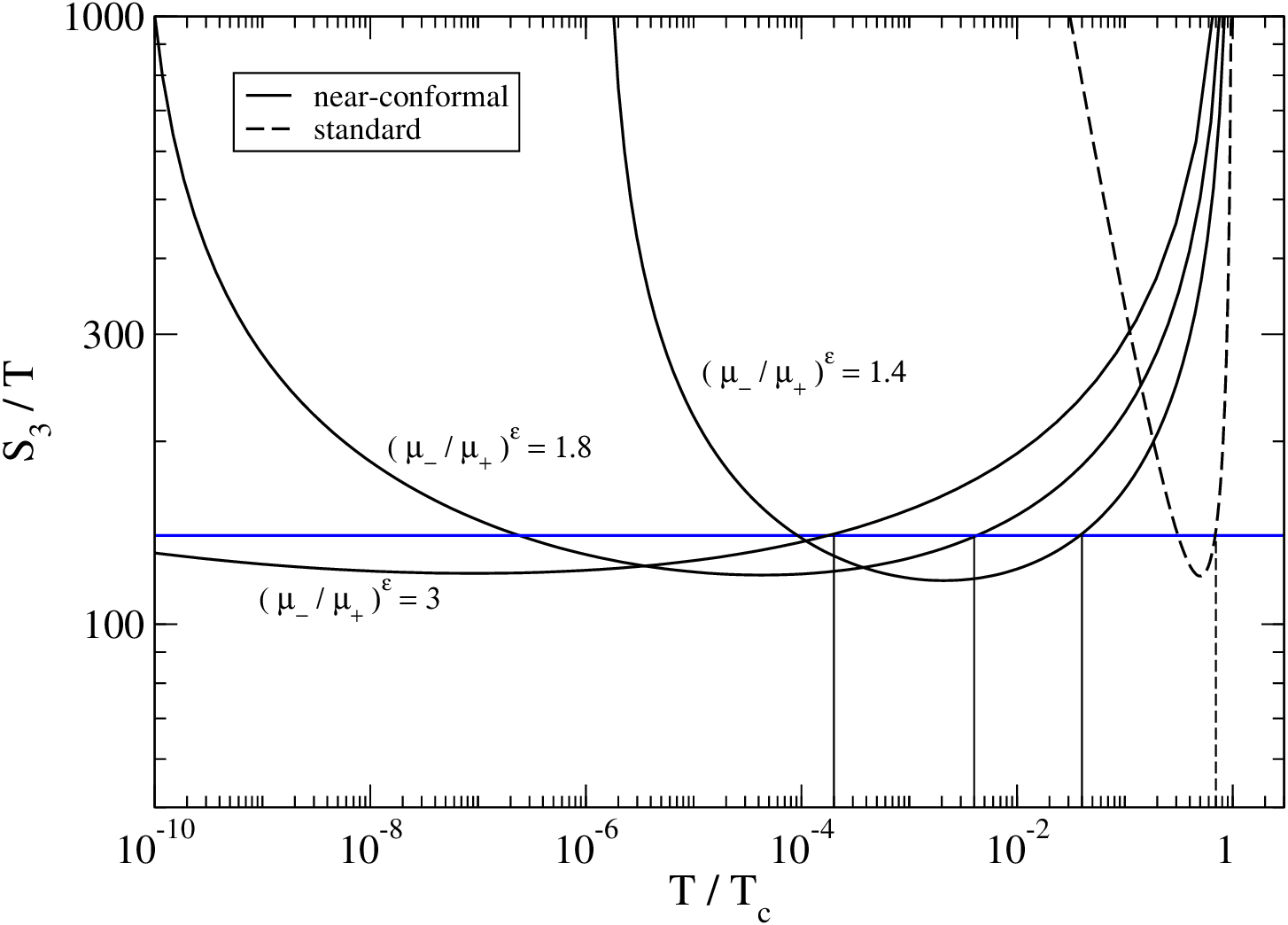}
\caption{
\label{fig:actionS3}  
\small The tunneling action $S_3/T$ as a function of $T/T_c$ for a
typical nearly conformal potential (solid line) (we used the
Goldberger-Wise potential for illustration) and for a usual polynomial
Higgs potential (dashed line). The horizontal blue line indicates the
tunneling value $S_3/T\sim 4 \log (M_{\rm Pl}/T_{\rm EW})\sim 140$.
For a standard potential, the nucleation temperature $T_n$ is always
close to the critical one, $T_c$, unless some fine-tuning is invoked.
For a nearly conformal potential, supercooling is a general feature and
$T_n$ can easily be several orders of magnitude below $T_c$.}
\end{center}
\end{figure}

In contrast, the potential (\ref{equation:nearlyconformal}) leads to a
tunneling action that is well-behaved as a function of $\mu^\epsilon$
rather than $\mu$. This way it is possible to achieve a small
nucleation temperature together with bubble percolation and a rather
long but finite duration of the phase transition for $\epsilon \sim
{\cal O}(1/10)$
\be \beta / H =\left.  T \frac{d}{dT} \frac{S_3}{T}
\right|_{T_n} \sim \epsilon \ \left.  \frac{S_3}{T} \right|_{T_n}
\gtrsim 1.  \label{eq:percolation} \ee 
An example is given in
Fig.~\ref{fig:actionS3} where the tunneling action is plotted for a
specific Goldberger-Wise potential~\cite{Goldberger:1999uk} (taken
from Ref.~\cite{Konstandin:2010cd}) in comparison with an action
occurring e.g.~in the electroweak phase transition in supersymmetric
extensions of the SM.

Let us explain this more quantitatively.  The conformal phase
transition can be studied by working in a five-dimensional Anti de
Sitter (AdS) space in which the radion is stabilized by a bulk scalar
with a relatively small mass \cite{Creminelli:2001th, Randall:2006py,
  Nardini:2007me, Konstandin:2010cd}.  In the 4D picture, this
corresponds to a balance between a marginal and a slightly irrelevant
deformation of the gluon sector of the CFT.  At high temperature, the
system is in an AdS-Schwarzschild (AdS-S) phase involving a single
ultraviolet (Planck) brane, providing the UV cutoff of the theory.
The free energy of the AdS-S phase is given by
\be
F_{\rm AdS-S}=-4 \pi^4 (Ml)^3 T^4,
\label{eq:ads}
\ee
where $l$ is the 5D AdS curvature, of the order of the Planck scale,
and $M$ the 5D Planck mass.  By holography, $F_{\rm AdS-S}$ can be
interpreted as the free energy of a strongly coupled large $N$ CFT
where the rank $N$ of the $SU(N)$ dual gauge theory in four dimensions
is related to $(Ml)^3$ via the AdS/CFT correspondence
\be
\label{N_ML3_relation}
{N^2}={16 \pi^2} (Ml)^3.
\ee
Note that this relation (\ref{N_ML3_relation}) holds for the
gauge/gravity duality in the ${\cal N}=4$ supersymmetric context. In a
more bottom-up approach to holography (like e.g.~in AdS/QCD) the four
dimensional theory has less supersymmetry and (\ref{N_ML3_relation})
is modified.

At low temperature, there are two branes, with a slice of AdS bulk in
between.  The infrared brane spontaneously breaks the conformal
symmetry of the theory.  The resulting effective potential of the
radion is of the form (\ref{equation:nearlyconformal}) where the field
$\mu$ is a reparametrization of the brane separation $r$
\be
\mu = l^{-1} e^{-r/l},
\ee
with a canonical kinetic term (up to a factor $12(M l)^3$,
see~\cite{Konstandin:2010cd}). The position of the extrema $\mu_\pm$
of $V$ depend on the specific parameters but are given by
\be
\mu_+^\epsilon \lesssim \mu_-^\epsilon \lesssim  1.
\ee
In the Randall-Sundrum model, the smallness of $\epsilon \sim
{\cal{O}}(1/10)$ is used to generate the hierarchy between the Planck
and the electroweak scale, $\mu_- \ll l^{-1}$, but also implies $\mu_+
\ll \mu_-$ and the potential is nearly conformal between those widely
spread values.

The tunneling action can be calculated by determining the bounce
solution for the potential (\ref{equation:nearlyconformal})
\cite{Creminelli:2001th,Randall:2006py}.  An accurate approximation
can be obtained by exploiting the nearly conformal behavior of the
system (we follow the notation and analysis
of~\cite{Konstandin:2010cd}). For a certain bounce solution with
release point $\mu_r$, the potential is approximated by
\be
\label{eq:app_conf}
V(\mu) \approx \mu^4 P((\mu_r/\mu_0)^\epsilon) \equiv - \mu^4 \kappa.
\ee
The conformal invariance of the potential then allows to determine the
action and the corresponding nucleation temperature $T_n$ as (we only
consider the $O(3)$ symmetric tunnel action here)
\be
\label{eq:est_conf}
S_3/T \simeq 290 \kappa^{-3/4} (M l)^3, \quad
T_n \simeq 0.1 \kappa^{1/4} \mu_r.
\ee
The action of a critical bubble is minimal when the quartic coupling
$\kappa$ is maximal.  Quite generally, the value of $\kappa$ is
bounded by a value around $\frac12$ ~\cite{Konstandin:2010cd}. Hence,
for large values of $(M l)^3$, the phase transition is strong and the
symmetric phase can even become stable~\cite{Creminelli:2001th}.
Accordingly, for the phase transition to complete, we must have
\be
\label{eq:ml_bound}
(M l)^3 \lesssim 0.3 \, .
\ee
We will display more precise conditions on the parameter space in the
next section. 
\begin{figure}[t!]
\begin{center}
\includegraphics[width=0.625\textwidth, clip ]{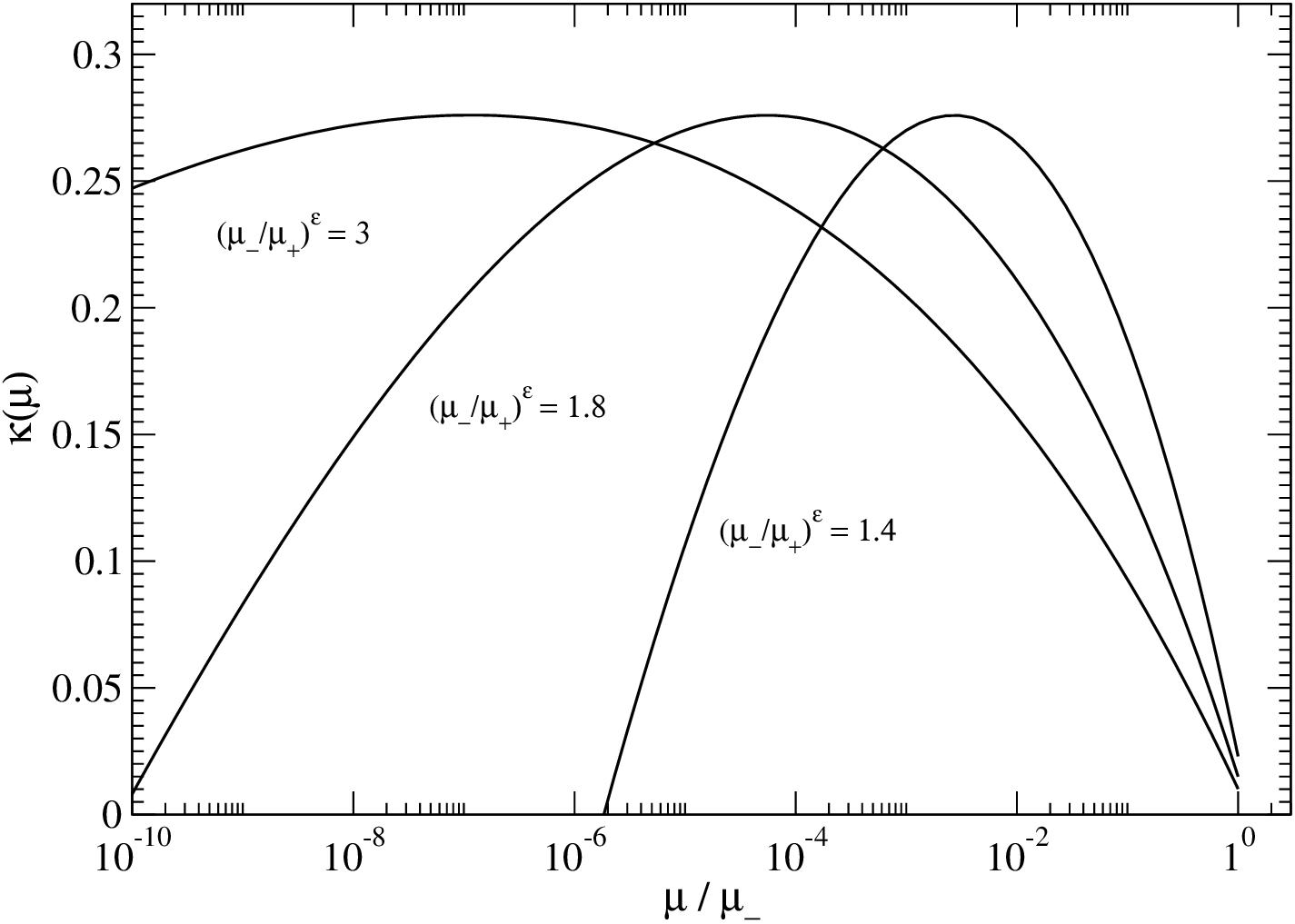}
\caption{
\label{fig:kappa} \small 
 The running of the effective quartic coupling
$\kappa$. Tunneling typically occurs when the quartic coupling is
close to maximal, for a value of the release point close to $\mu_r
\gtrsim \sqrt{\mu_+ \mu_-}$, which can be orders of magnitude smaller
than the value $\mu_-$ at the minimum of the potential.}
\end{center}
\end{figure}

The function $S_3/T$ is shown in Fig.~\ref{fig:actionS3} and $\kappa
(\mu)$ is plotted in Fig.~\ref{fig:kappa}.  For most of the parameter
space, the system tends to tunnel close to the maximum of $\kappa$.
Since $\kappa$ is a second order polynomial function of $\log \mu$ in
the limit $\epsilon \ll 1$, its maximum is given by $\log \mu_r
\gtrsim (\log \mu_+ + \log \mu_-)/2$, hence
\be
\label{eq:murbound}
\mu_r \gtrsim \sqrt{\mu_- \mu_+} \ll \mu_-, \quad T_n \ll T_c. 
\ee
Therefore, the Universe tends to be cold at the onset of the phase
transition. On the other hand, for the duration of the phase transition
one finds
\bea
\beta / H &=& \left.  T \frac{d}{dT} \frac{S_3}{T} \right|_{T_n} 
\approx \frac{3}{4}\frac{S_3}{T} 
\left( \mu_r \frac{d}{d\mu_r} \kappa \right) \sim \frac{3 \epsilon}{4}\frac{S_3}{T},  
\eea
and percolation is not an issue for $\epsilon \gtrsim 10^{-2}$.

Next, we investigate what are the typical expectations for the number
of efolds and what controls the amount of supercooling. For that, we
use the explicit Goldberger-Wise potential.

\subsection{Typical amount of supercooling (number of efolds)}

The number of efolds of inflation is given by
\be
N_{\rm efolds } \sim  \log \frac{T_c}{T_n}.
\ee
In this section we estimate the two temperatures $T_c$ and $T_n$ and
the typical amount of supercooling of the conformal phase transition.
As we have seen previously, the nucleation temperature $T_n$ is
proportional to the release point $\mu_r$ :
\be
T_n \simeq 0.1 \,  \mu_r \, \kappa(\mu_r)^{1/4},
\ee
which is bound by the position of the maximum of the quartic coupling
$\kappa(\mu)$ given by the scale $\sqrt{\mu_- \mu_+}$. In the
Goldberger-Wise mechanism of the Randall-Sundrum model, parameters of
the potential are fixed such that the ratio $\mu_-/\mu_+$ explains the
hierarchy between the EW scale and the Planck scale.  Following the
notations used in \cite{Konstandin:2010cd}, the ratio $\mu_-/\mu_+$
can be written as
\be
\frac{\mu_-}{\mu_+} = \lp \frac{\xi_-}{\xi_+} \rp^{1/\epsilon} 
\ \ \mbox{where} \ \  \xi = \frac{v_1}{v_2} e^{- \epsilon r / l},
\ee
and $v_1$ and $v_2$ are the expectation values of the Goldberger-Wise
bulk scalar field at the two boundaries.  The parameters $\xi_-$,
$\xi_+$ and $\epsilon$ are related by (see eq.~(49) of
\cite{Konstandin:2010cd})
\begin{eqnarray}
\label{eq:xiplus}
\xi_+ &=& \frac{4 + \epsilon}{2+  \epsilon} - \xi_- \approx 2 - \xi_- 
\ \ \mbox{for small detuning on the brane} \\
\xi_+ &\approx &\xi_-^2 \frac{v_2}{v_1} \ \ 
\mbox{ for large detuning on the brane}
\label{eq:ximinus}
\end{eqnarray}
The detuning is to a modification of the values of the two brane
tensions that are adjusted to ensure a vanishing 4D cosmological
constant in the absence of back reactions. In particular one always
has
\be
\label{eq:xi_bound}
\xi_+ > \xi_-^2 \frac{v_2}{v_1}.
\ee
We therefore obtain from the EW/Planck scale hierarchy  the upper bound
\be
\frac{\mu_-}{\mu_+} = \lp \frac{\xi_-}{\xi_+} \rp^{1/\epsilon} 
< e^{r_-/l} \approx e^{37} \approx  10^{16},
\label{eq:hierarchy}
\ee
which in turn gives a lower bound on the nucleation temperature
according to (\ref{eq:murbound}).
\begin{figure}[ht]
\begin{center}
\includegraphics[width=0.475\textwidth, clip ]{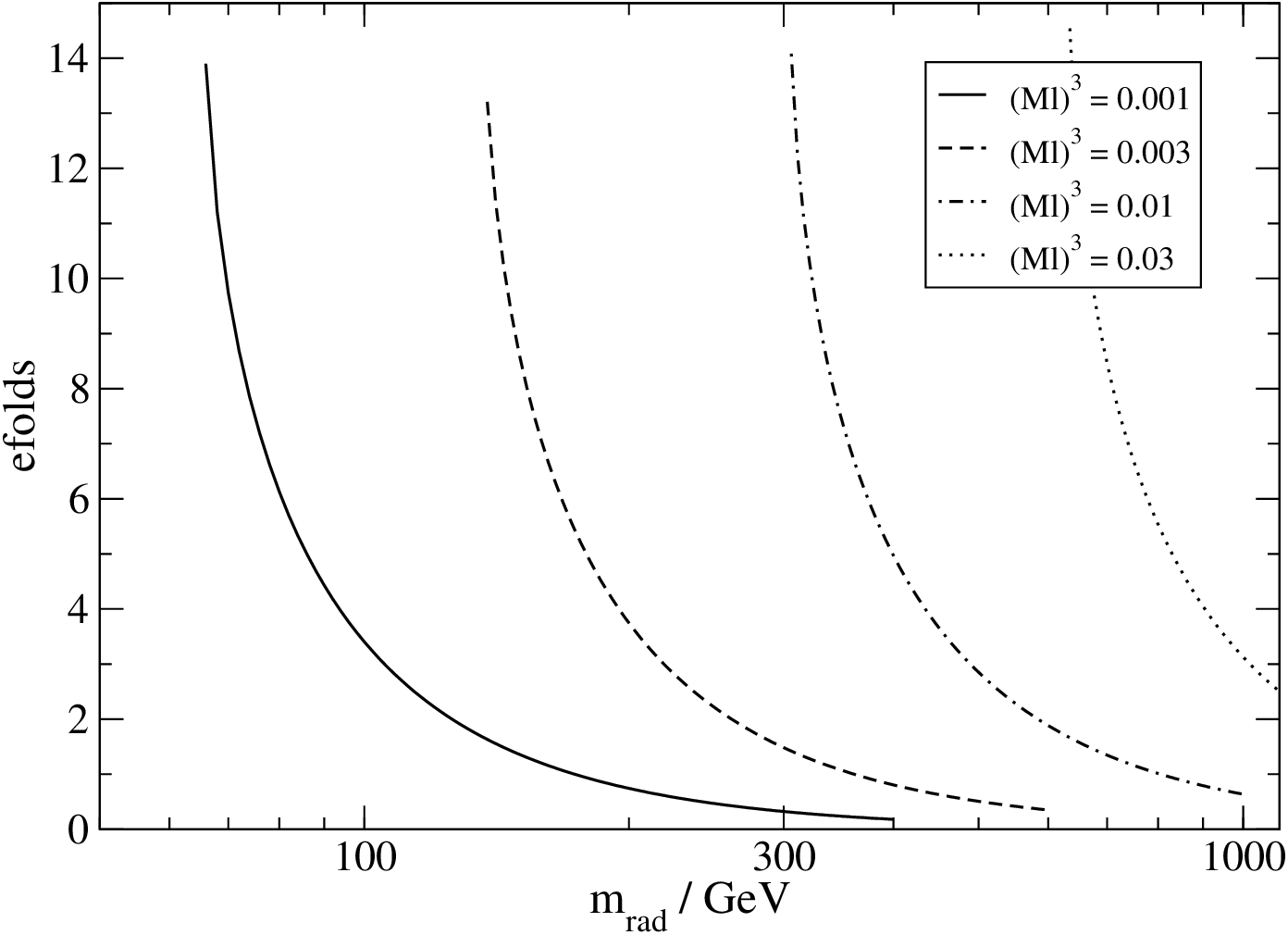}
\includegraphics[width=0.475\textwidth, clip ]{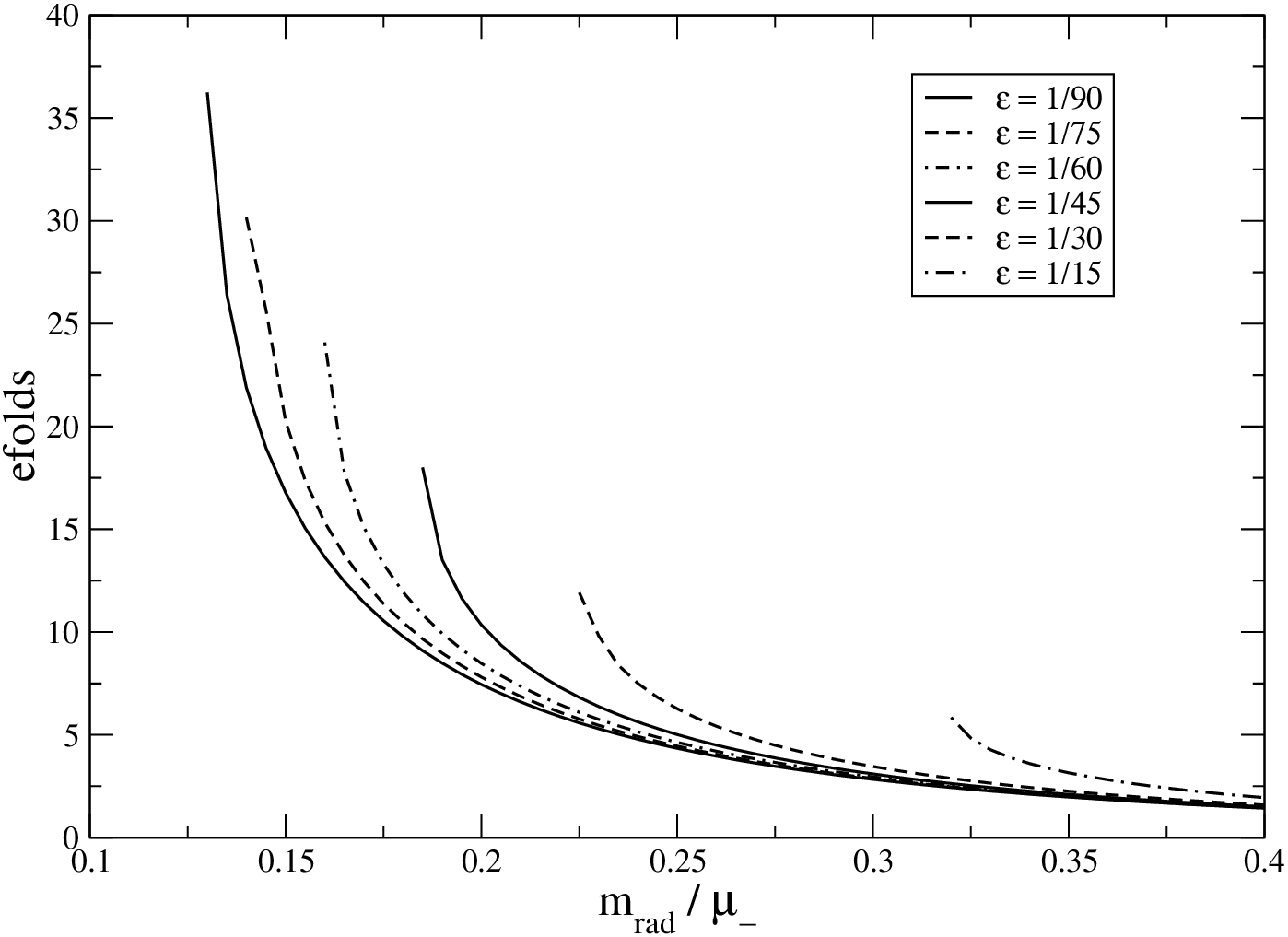}
\caption{
\label{fig:efolds}
\small Number of efolds of inflation as a function of the radion mass.
Left: Randall-Sundrum model for different values of $(Ml)^3$ and
$\mu_- = 4$ TeV; Right: A generic model with potential
(\ref{equation:nearlyconformal}) where the constraint fixing the
hierarchy, eq.~(\ref{eq:hierarchy}), is relaxed. At the point where
the curves stop, the system cannot tunnel and is stuck in the
symmetric phase.}
\end{center}
\end{figure}

As for the critical temperature $T_c$, it is given by equating the
free energy of the AdS-S phase, with the potential
difference between the conformally symmetric and broken phases. This
potential difference is given by
\be
\Delta V = 2 \epsilon l^{-4} (M l )^3 \frac{v_2^2}{M^3} 
\xi_- (\xi_- -1) e^{-4 r_-/l}, 
\ee
and using the expression for the radion mass (eq.~(69) of
\cite{Konstandin:2010cd})
\be
\label{eq:mrad}
m^2_{\rm rad} = \frac23 \epsilon \mu_-^2 \frac{v_2^2}{M^3} 
(4(\xi_- - 1) + \epsilon \xi_- ) \xi_-,
\ee
we have for $\xi_-$ not too close to unity
\be
\Delta V \approx \frac34 (M l)^3 m_{\rm rad}^2 \mu^2_-.
\ee
Equating this with the free energy of the AdS-S phase,
eq.~(\ref{eq:ads}), yields
\be
4 \pi^4 T_c^4 = \frac34 m_{\rm rad}^2 \mu_-^2,
\ee
leading to
\be
\label{eq:Nefolds}
N_{\rm efolds} \sim \log \frac{T_c}{T_n}
\simeq \log \frac{\mu_-}{\mu_r} + 0.74 - 
\frac14 \log \kappa(\mu_r) + \frac12 \log  m_{\rm rad}/ \mu_-.
\ee
Therefore, the number of efolds is essentially controlled by the
hierarchy $\mu_-/\mu_+$ and the maximal number of efolds is 
\be
N_{\rm efolds} < \log \sqrt{\frac{\mu_-}{\mu_+}} = \frac{r_-}{2 l} \approx 18.
\ee
We plot in Fig.~\ref{fig:efolds}(a) the number of efolds (more
precisely $\log \mu_-/\mu_r$) as a function of the radion mass in the
Randall-Sundrum model where the ratio $\mu_-/\mu_+$ is constrained by
the weak/Planck scale hierarchy.  This constraint is relaxed in
Fig.~\ref{fig:efolds}(b) where therefore the number of efolds can
reach values required to solve the horizon problem. In any case, we
see that a large number of efolds is associated with a light radion
(relative to the scale $\mu_-$).


\subsection{Backreaction constraints}

In this section, we derive the limits to the validity of our analysis,
in particular the constraints from backreaction.  In the
Goldberger-Wise stabilization mechanism \cite{Goldberger:1999uk} that
leads to a potential of the form (\ref{equation:nearlyconformal}), one
introduces a 5D bulk scalar $\Phi $ with a mass $m$ that is related to
$\epsilon$ by 
\be 
\epsilon = \sqrt{4 + m^2 l^2} - 2.  
\ee 
As discussed in \cite{Randall:2006py}, imposing that the energy in the
Goldberger-Wise field is subdominant compared to the bulk cosmological
constant (in other words that it does not distort too much the AdS
geometry) significantly restricts the available parameter space.

\begin{figure}[t!]
\includegraphics[width=0.3295\textwidth, clip ]{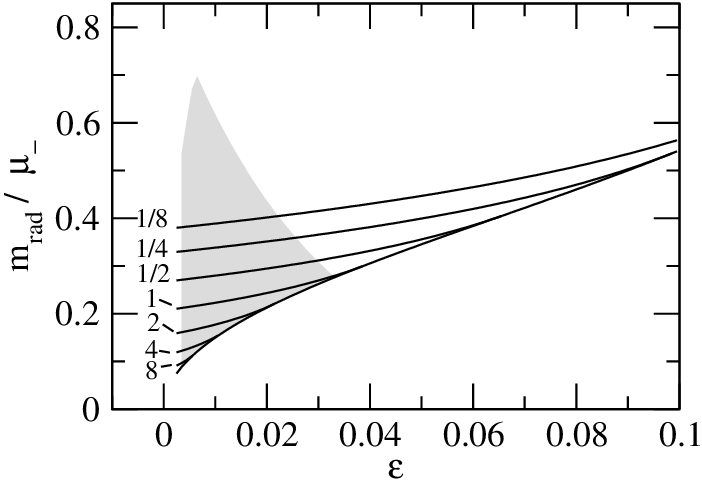}
\includegraphics[width=0.3295\textwidth, clip ]{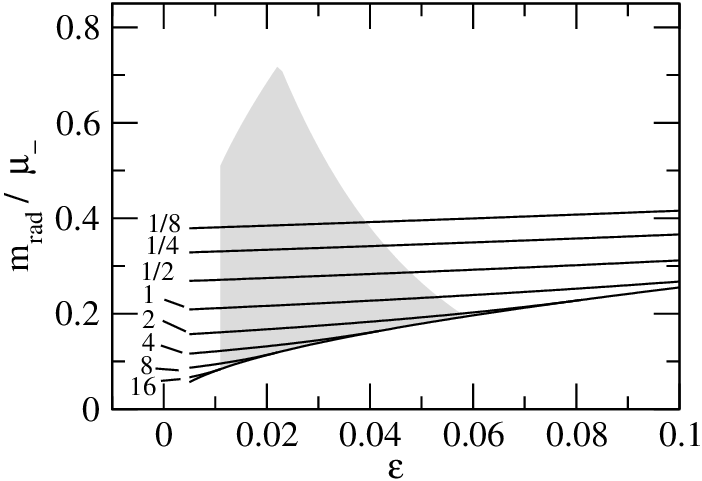}
\includegraphics[width=0.3295\textwidth, clip ]{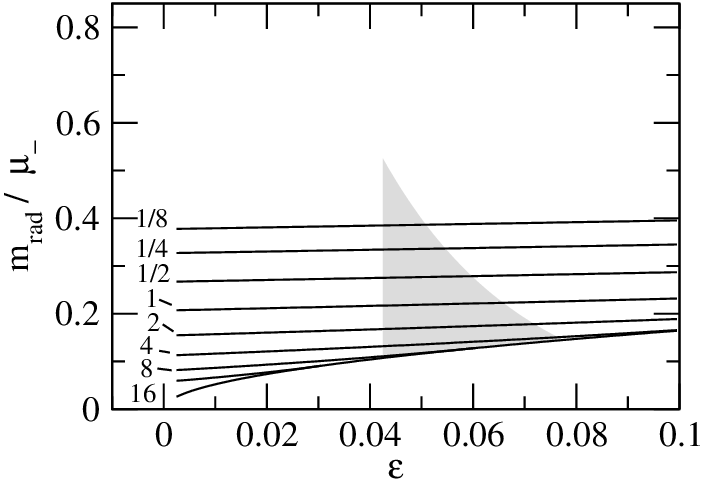}
\includegraphics[width=0.3295\textwidth, clip ]{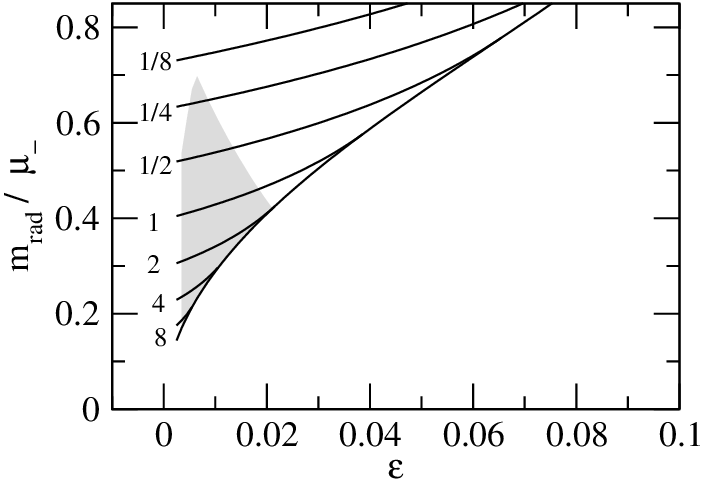}
\includegraphics[width=0.3295\textwidth, clip ]{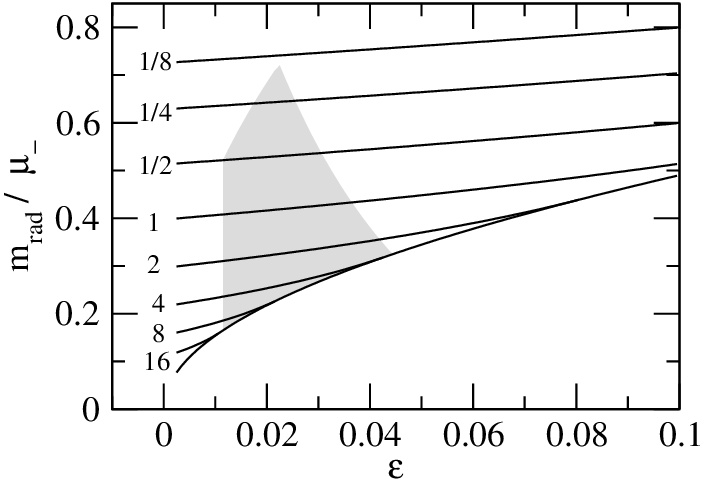}
\includegraphics[width=0.3295\textwidth, clip ]{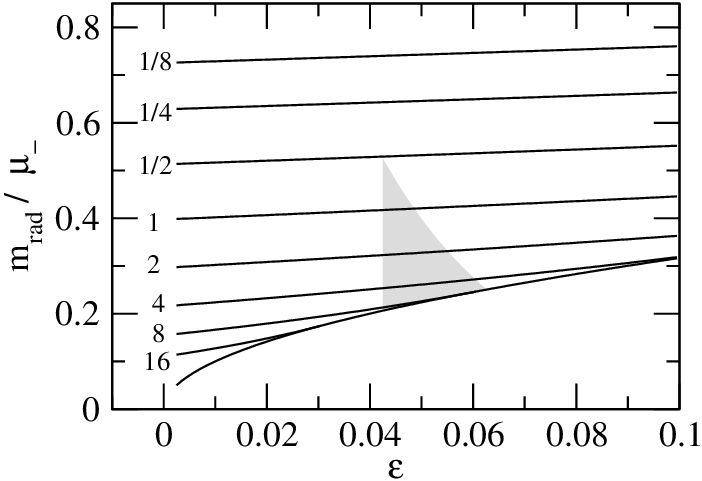}
\includegraphics[width=0.3295\textwidth, clip ]{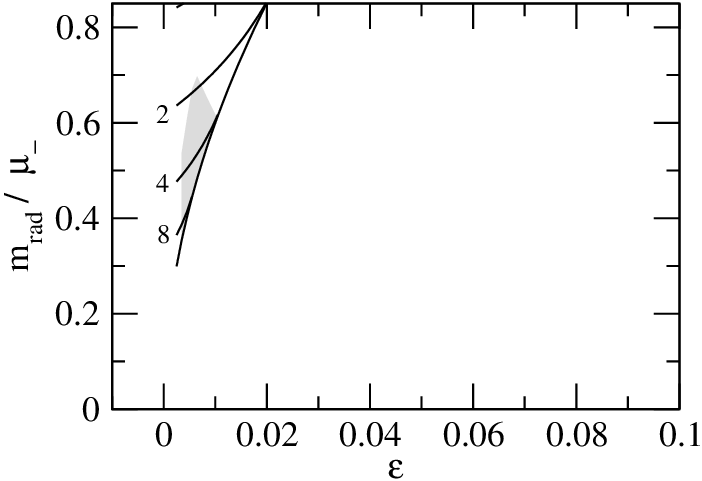}
\includegraphics[width=0.3295\textwidth, clip ]{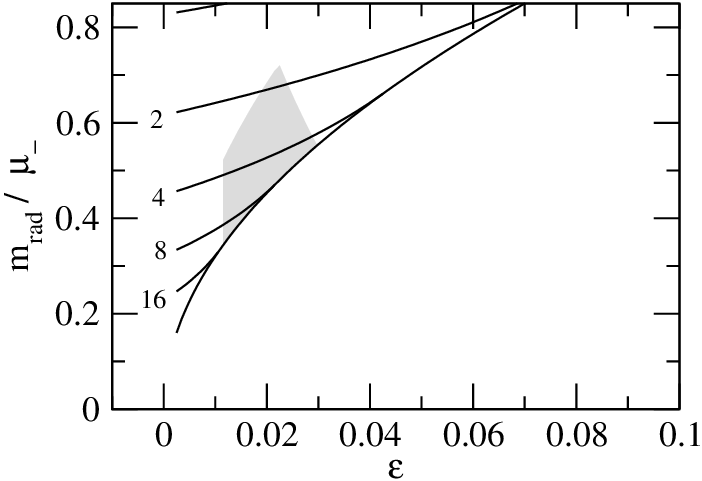}
\includegraphics[width=0.3295\textwidth, clip ]{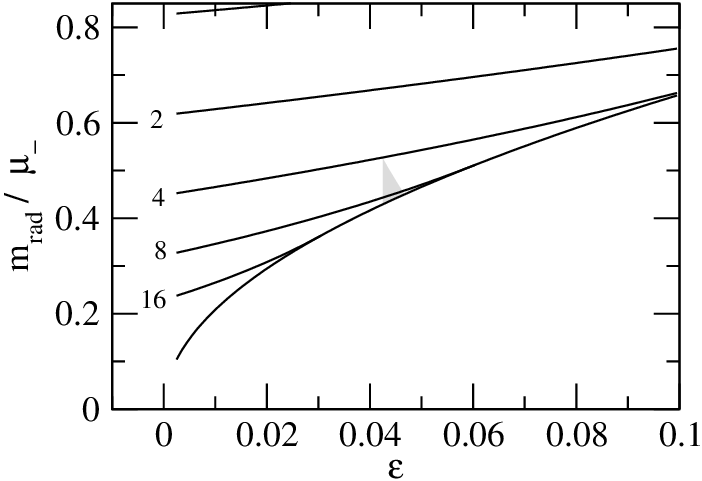}
\caption{
\label{fig:eContours}
\small Contours for the number of efolds (more precisely $\log
\mu_-/\mu_r$). The shaded region is where calculability can be
trusted, as defined by the constraints in
(\ref{eq:efolds_constraint}). Below the bottom line, the system never
tunnels to the broken phase. From top to bottom, the plots show $N=2$,
$3$ and $5$; from left to right the series of three plots respectively
use $\xi_-/\xi_+=1.05$, $1.2$ and $1.65$. For larger $N$ the phase
transition becomes stronger and beyond $N>6$ the system is generally
stuck in the symmetric phase, at least in the domain of calculability
and only considering thermal tunneling.  }
\end{figure}

The 5D metric is parametrized as
\be
ds^2 = e^{2 A(r)} \left( dt^2 - e^{2 \sqrt{\Lambda} t} d\vec{x}^2 \right) - dr^2,
\ee
where $\Lambda$ is the 4D cosmological constant.  Our parameter space
comprises the radion mass $m_{rad}$ , $\epsilon$ and $(Ml)^3$ (or $N$
in the 4D language) and is constrained by requiring that some terms in
the equation of motion of the warp factor $A$ are small:
\bea
A^{\prime 2}&=& \frac{1}{l^2} - \frac{1}{24 M^3} m^2 \Phi^2 + \frac{1}{24 M^3} \Phi^{\prime2} 
+ \Lambda e^{-2A}. \label{eq:A_EoM}
\eea
Demanding that the first term in (\ref{eq:A_EoM}) dominates over the
second and third leads to
\be
4 \epsilon v_1^2 \ll 24 M^3, \quad {\rm and } \quad 
v_2^2 ( \epsilon \xi_- + 4( \xi_- - 1))^2 \ll 24 M^3.
\ee
In addition, we should consider the impact of the 4D
cosmological constant term $\Lambda$ on the parameter space. 
If it is neglected, the parameters $\xi_-$ and $\xi_+$ are
related by (\ref{eq:xiplus}) while in the regime of a large
cosmological constant one finds the relation (\ref{eq:ximinus})
and in general the bound (\ref{eq:xi_bound}).  Interestingly, this
bound automatically ensures percolation,
c.f.~eq.~(\ref{eq:percolation}).
Using the radion mass (\ref{eq:mrad}) and the relation $\xi_- =
\frac{v_1}{v_2} e^{-\epsilon r_-/l}$, the three constraints are
\bea
\label{eq:efolds_constraint}
m^2_{\rm rad}/\mu_-^2 &<& 16 \epsilon \frac{\xi_-}{4(\xi_- - 1) + \epsilon \xi_-}, \nn \\
m^2_{\rm rad}/\mu_-^2 &<& 4 e^{-2 \epsilon r_-/l}
\frac{4(\xi_- - 1) + \epsilon \xi_-}{\xi_-}, \nn  \\
\epsilon &>& \frac{l}{r_-} \log \xi_- / \xi_+.
\eea
These constraints together with the contour lines for the predicted
number of efolds are shown in Fig.~\ref{fig:eContours} for three
values of $(Ml)^3= {N^2}/{16 \pi^2}$.  These plots clearly demonstrate
the observation first made in Ref.~\cite{Creminelli:2001th} that there
is a tension between the large $N$ assumption needed for calculability
and the possibility to complete the phase transition.  In
\cite{Creminelli:2001th}, the bounds were stronger and the conclusion
was rather negative, i.e that the transition could not complete in the
regime of calculability.  This conclusion was ameliorated in
\cite{Randall:2006py} where the tunneling action was estimated in the
supercooling regime, namely in the thick-wall limit, and for $O(4) $
symmetric bubbles and also taking into account the fact that the field
value at tunneling is not close to the value at the minimum. These
effects improve the nucleation probability, as re-examined in more
details in \cite{Nardini:2007me}, and refined by taking into account
backreactions in \cite{Konstandin:2010cd}, confirming that the bounds
are actually less stringent than in \cite{Creminelli:2001th}.  One
also gets a much weaker phase transition when the geometry is deformed
in the infrared~\cite{Hassanain:2007js}.
%

To conclude, in the region of parameter space allowing calculability,
the phase transition tends to be so strong that several efolds of
inflation can occur before the onset of the phase transition. If one
is willing to push calculability to its limit, $N=2$, then there is
typically less supercooling and several efolds take place only by
tuning the radion mass to a low value.  Finally, note that in this
paper, we present results only for $\epsilon>0$.  It would be
interesting to consider in more details the $\epsilon<0$ case which
was argued to be more promising in \cite{Randall:2006py}, although in
this case strong coupling effects arise in the IR and may be important
before nucleation, thus preventing a reliable analysis.
Furthermore, our discussion has been based on the Goldberger-Wise
 potential.  There
are alternative stabilization mechanisms relying on the Casimir forces
induced by bulk fields \cite{Garriga:2002vf}. Although we do not
expect that the picture we have exposed would be qualitatively
different, it would certainly be interesting to study the phase
transition in this context.

\subsection{Graceful first-order inflation from nearly conformal dynamics }

Guth's original idea of inflation \cite{Guth:1980zm} was precisely
that inflation could end by the tunneling of the false vacuum into the
true vacuum during a first-order phase transition. However, it was
realized that true vacuum bubbles would not percolate to give rise to
the primordial plasma of relativistic degrees of freedom
\cite{Guth:1982pn}. This drawback was solved in slow-roll models of
inflation \cite{Linde:1981mu, Albrecht:1982wi} where a scalar field,
the inflaton, slowly rolls down along its very flat potential.

As shown in the previous section, with a nearly conformal potential
(\ref{equation:nearlyconformal}) it is possible to have a stage of
inflation ended by a first-order phase transition, with no particular
fine-tuning in the potential. However, it is clear from fig
\ref{fig:efolds}(a) that we cannot reach naturally a sufficiently
large number of efolds to solve the standard cosmological problems
that inflation is supposed to solve, even though for an inflationary
stage taking place at the electroweak scale, we need only $\sim 30$
efolds to solve the horizon problem rather than $\sim 60$ if the
inflation scale $M_I$ is at the GUT scale, according to
\be N_{\rm efolds}
= 62 - \log \left( \frac{10^{16} \, {\rm GeV}} {M_I} \right) - \frac{1}{3} \log
\left(\frac{M_I}{T_{\rm reh}}\right). 
\ee
Besides, even if we tune the radion mass so that a large number of
efolds is achieved, one still has to solve the problem of generation
of density perturbations as reheating from bubble collisions can only
produce isocurvature density perturbations \cite{Turner:1992tz} and an
extra source is needed to explain the generation of density
perturbations.  Anyhow, we note that there are no constraints on our
extended stage of supercooling from either Big Bang Nucleosynthesis or
from the Cosmic Microwave Background \cite{Turner:1992tz}.  Imposing
that there is no large fraction of bubbles that are Hubble size when
nucleosynthesis commences and similarly that there is no large
inhomogeneous regions near the last scattering surface that would lead
to distortions in the CMB, results in weaker conditions than our
criterion for percolation (\ref{eq:percolation}).

\section{Reheating temperature and implications for baryogenesis and dark matter}
\label{sec:reh}

Phenomenological consequences of a strongly first-order phase
transition have been extensively discussed in the literature.  We
stress here that we distinguish the phase transition associated with
conformal symmetry breaking from the one associated purely with EW
symmetry breaking, although they are intertwined. The full EW symmetry
breaking sector has a potential of the form
\be V_{\rm TOT}= \mu^4 \times \left( \
P(({\mu}/{\mu_0})^{\epsilon})+ {{\cal V} (\phi)}/{\mu_0^4} \ \right).
\ee 
While $\mu$ condensation induces EW symmetry breaking (and thus
bubbles also involve Higgs field varying vev), one should keep in mind
that the potential ${\cal V} (\phi) $ alone may not necessarily
display a first-order phase transition.  On the other hand, it is the
phase transition associated with $\phi$ condensation which is
traditionally the relevant one for baryogenesis.  The nature of the
phase transition in composite Higgs models remains to be investigated
in specific models. Studies of the scalar potential have concentrated
on ${{\cal V} (\phi)}$ \cite{Agashe:2004rs, Gripaios:2009pe}. However, one
should in principle compute the full $V_{\rm TOT}(\mu,\phi)$, which is
a non-trivial task.  Although this is a model-dependent question that
relies on the form of the Higgs potential ${\cal V} (\phi) $ which we
do not specify here, some general statements can be made on the
cosmology as we discuss now.

\subsection{Reheating temperature predictions}

In our scenario, when bubbles are nucleated, the universe is very cold.
Reheating starts when bubble collide.
The value of the reheat temperature is a crucial ingredient to
determine not only what are the possible frameworks for baryogenesis
but also what are the underlying conditions for the computation of the
dark matter abundance. Baryogenesis depends on whether the reheat
temperature $T_{\rm reh}$ is below or above the sphaleron
freeze-out temperature, which is essentially given by the temperature
at which the electroweak symmetry is broken. As far as dark matter
is concerned, if $T_{\rm reh}$ is at the electroweak scale,
the common thermal freeze-out mechanism prediction for WIMP dark
matter abundance is no more guaranteed if $T_{\rm reh}
\lesssim m_{\rm DM} $.

The process of reheating from bubble collisions was first discussed
in~\cite{Hawking:1982ga} and later in the early nineties in 
\cite{Watkins:1991zt, Kibble:1995aa, Kolb:1996jr}. We apply these
results to the case of a strongly first-order phase transition taking
place in an empty universe as a result of nearly conformal dynamics at
the TeV scale in the companion paper~\cite{Konstandin_Servant2}, where, in
particular, we argue that the scalar field dynamics during the
collisions provides ideal conditions for a natural {\it cold }
baryogenesis mechanism. In the present section, we discuss the value
of the reheat temperature $T_{\rm reh}$ and review the consequences
for baryogenesis and dark matter generation mechanisms in this new
context.

At the TeV scale, the expansion of the Universe is negligible and
$T_{\rm reh}$ can be estimated using energy conservation
\be
\label{eq:low_reheat}
 \Delta V = g^*
\frac{\pi^2}{30} T_{\rm reh}^4,
\ee
where $g^* \sim 100$ is the number of relativistic degrees of freedom
given by the particle content of the SM after the phase transition. On
the other hand, the critical temperature
is\cite{Creminelli:2001th,Randall:2006py,Nardini:2007me,Konstandin:2010cd}
\be 
4 \pi^4 (Ml)^3 T_c^4 = \Delta V. 
\ee 
Note that in contrast with more common phase transitions, the
reheating temperature may therefore exceed the critical
temperature associated with conformal symmetry breaking, if there is a
large number of degrees of freedom in the CFT gas:
\be
(Ml)^3 \gtrsim \frac{g_*}{120 \pi^2}\sim 8 \times 10^{-2} 
\ \ \mbox{or} \ \ \  N > \sqrt{\frac{2 g_*}{15}}  \sim 3.6 \, .
\label{eq:Ml3bound}
\ee
The condition (\ref{eq:Ml3bound}) is only indicative. First of all,
after conformal symmetry breaking, most degrees of freedom of the
conformal sector have masses comparable to the temperature and will,
to a certain extent, contribute to the free energy, modifying the
relation (\ref{eq:low_reheat}).  Secondly, for a nearly conformal radion
potential, the tunneling back to the symmetric phase involves sizable
superheating and happens at a significantly larger temperature than
the critical one (just as the tunneling to the broken phase
involves sizable supercooling). In any case, even if nucleation back
to the symmetric phase is a priori possible, bubbles of symmetric
phase cannot grow rapidly into the broken phase, as the latent heat is
negative (see e.g.~\cite{Espinosa:2010hh}) and the phase transition has
to proceed by other means (e.g.~a slow growth of droplets). In
summary, we do not need to consider any particular constraint
resulting from a reheating temperature potentially higher than the
critical temperature associated with conformal symmetry breaking.

\begin{figure}[!t]
\begin{center}
\includegraphics[width=0.5\textwidth, clip ]{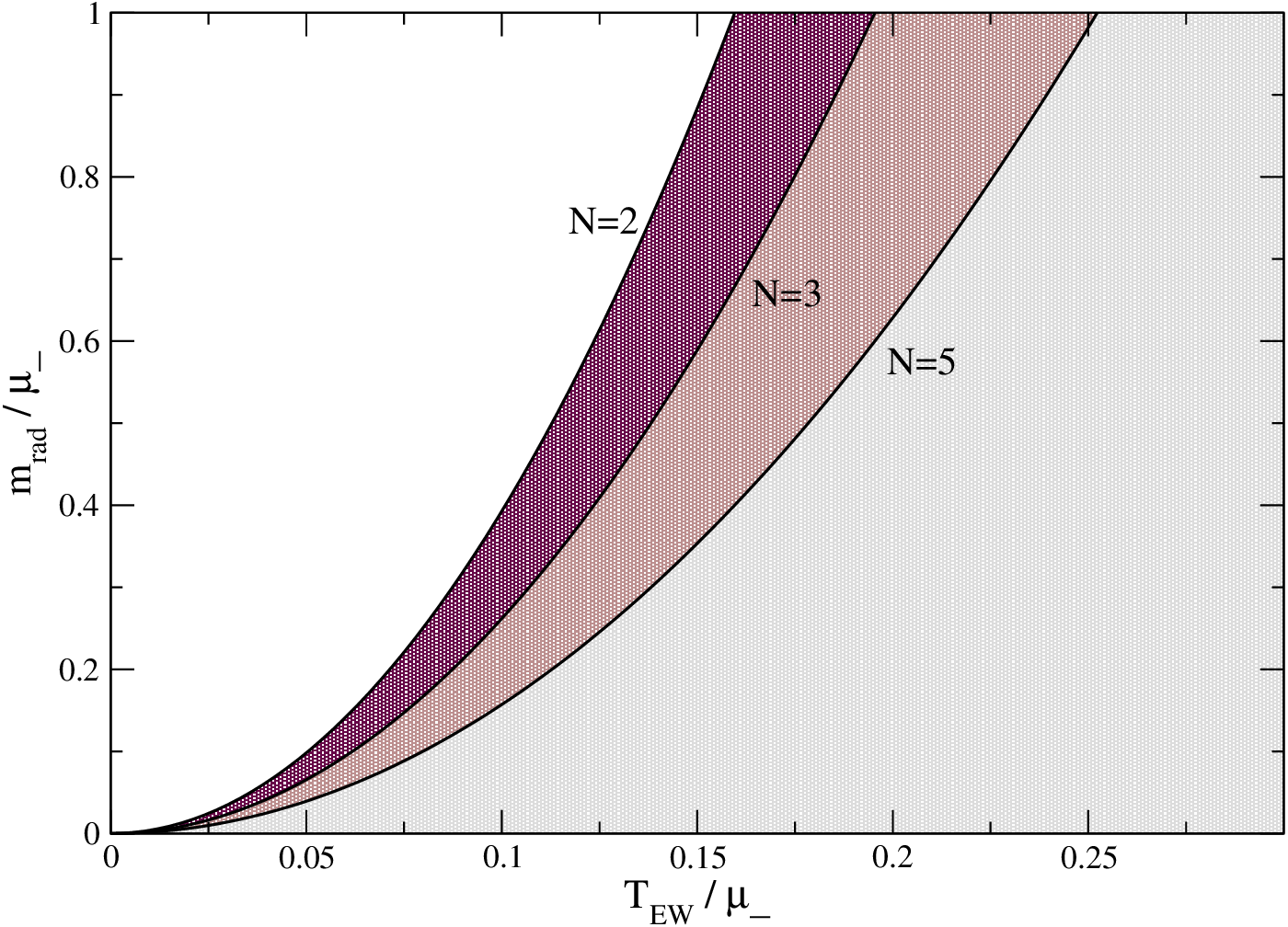}
\caption{\label{fig:reheat_constraint} { \small Value of the radion
mass (in the colored regions) for which the reheat temperature is
below the temperature at which the EW symmetry is restored,
$T_{\rm EW}$, according to eq.~(\ref{conditionTreheat}).}}
\end{center}
\end{figure}

\begin{figure}[t!]
\begin{center}
\includegraphics[width=0.475\textwidth, clip ]{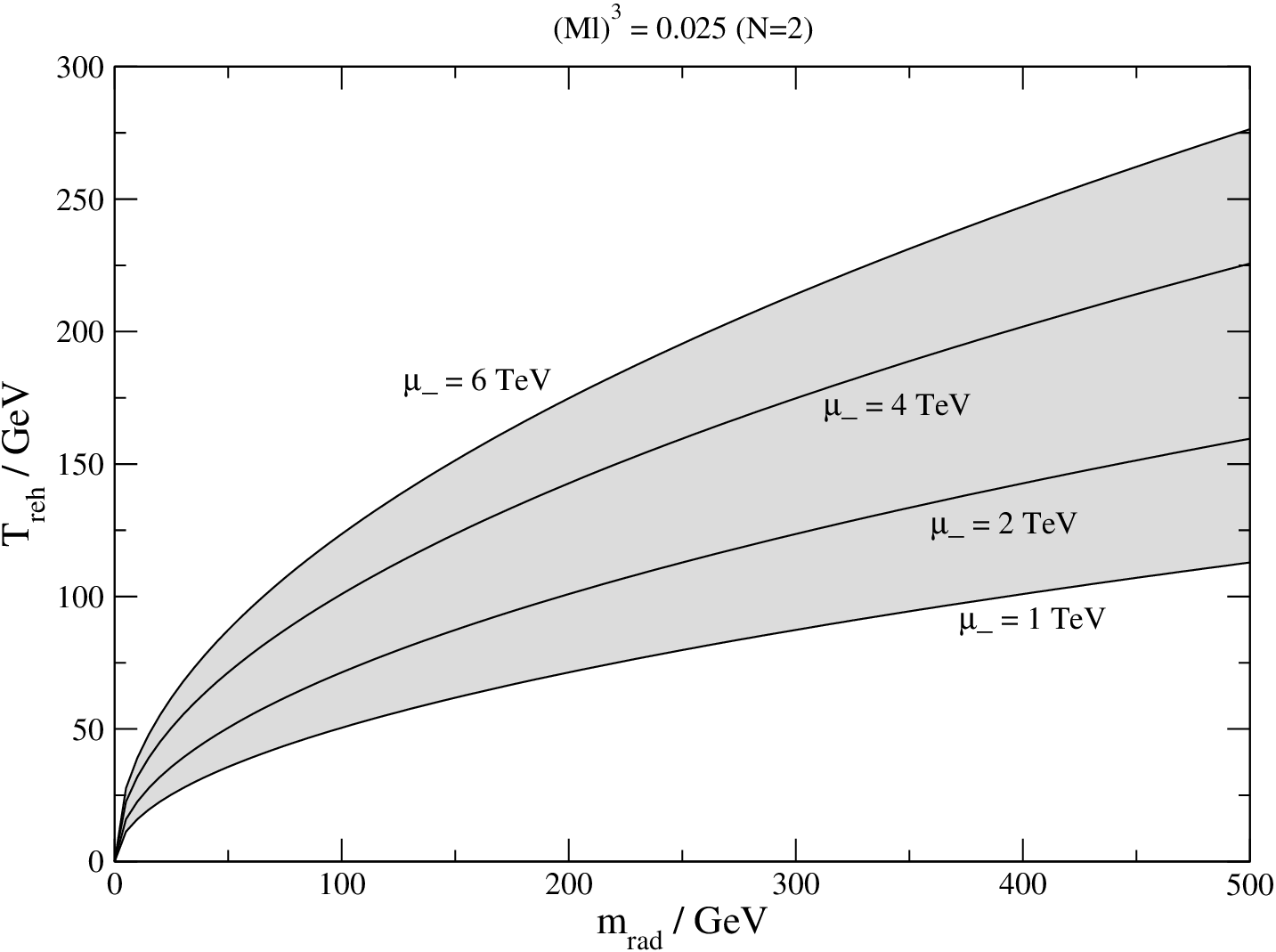}
\includegraphics[width=0.475\textwidth, clip ]{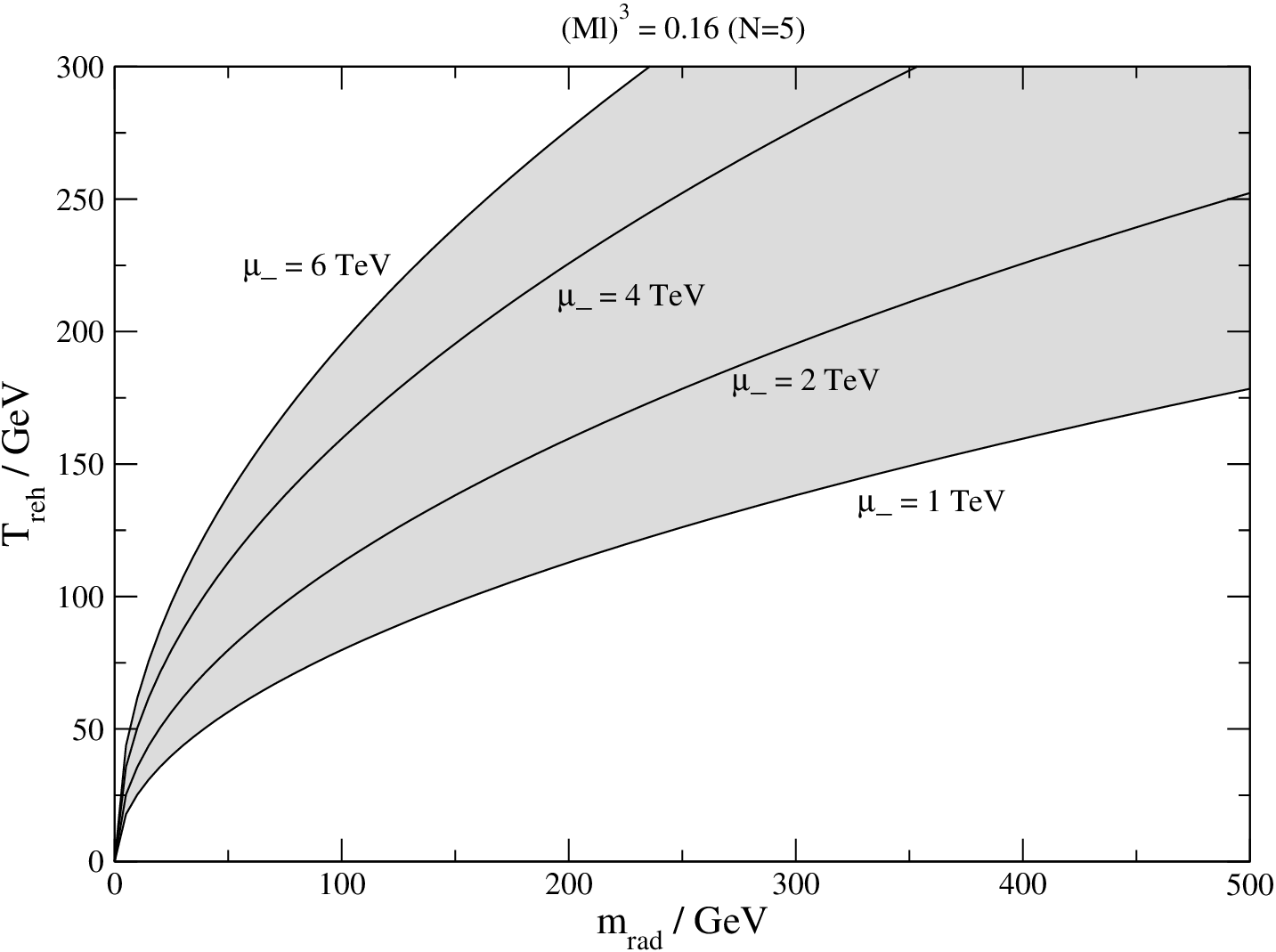}
\caption{\small Reheat temperature $T_{\rm reh}$ as a
function of the radion mass for different values of the scale of
conformal symmetry breaking $\mu_-$ and two values of $(Ml)^3$,
according to eq.~(\ref{conditionTreheat}).  }
\label{fig:Treheat}
\end{center}
\end{figure}

What we are really interested in is whether the electroweak symmetry
can be restored even if the conformal symmetry stays broken after
reheating.  Although this essentially depends on the form of the
effective potential for the Higgs field that we have not specified
here, some general conclusions can be drawn.  For instance, if the
reheat temperature is smaller than $T_{\rm EW}$, the temperature at which
the EW symmetry is restored, then standard EW baryogenesis cannot take
place and we have to rely on a different mechanism to explain the
matter-antimatter asymmetry of the universe.  Alternative mechanisms
exist and will depend on the reheating temperature that we estimate
now.  Using again the following expression for the free energy
difference in terms of the radion mass
\be
\Delta V = \frac34 (M l)^3 m^2_{\rm rad} \mu_-^2,
\ee
and (\ref{eq:low_reheat}) we obtain for the reheating temperature
\be
T_{\rm reh} = \left(\frac{45}{2\pi^2g_*}\right)^{1/4} (M l)^{3/4} 
\sqrt{m_{\rm rad} \mu_-},
\ee
which leads to
\be
T_{\rm reh} \gtrless T_{\rm EW} \ \ \longrightarrow \ \ 
\frac{m_{\rm rad}}{\mu_-} \gtrless \frac{6.6}{(Ml)^{3/2}}\left(
\frac{T_{\rm EW}}{\mu_-}\right)^2.
\label{conditionTreheat}
\ee
The bounds on ${m_{\rm rad}}/{\mu_-}$ in the Goldberger-Wise model
are shown in Fig.~\ref{fig:eContours} and the bound
(\ref{conditionTreheat}) is illustrated in
Fig.~\ref{fig:reheat_constraint}. We plot the reheat temperature as a
function of the radion mass in Fig.~\ref{fig:Treheat}. Thus, whether
the electroweak symmetry is restored after reheating depends on the
radion mass and on the temperature of electroweak symmetry breaking which
in turn depends on the Higgs mass.

For example, in the minimal composite Higgs model \cite{Agashe:2004rs},
rather large Higgs masses can arise, in particular at small $N$,  
while for the temperature of electroweak symmetry breaking, the Standard Model
relation~\cite{Anderson:1991zb} is valid
\be
\frac{T_{\rm EW}}{M_{\rm Higgs}} \simeq 
\left( m_W^2/v^2 + \frac12 m_Z^2 / v^2 + m_t^2 / v^2 \right)^{-1/2}
\sim 1.3 \, .
\ee
In the case $N=3$, significant
supercooling happens if $m_{\rm rad}/\mu_- < 0.3$ and for a Higgs mass of
200 GeV, the electroweak symmetry is restored after reheating if $\mu_-
> 2.6$ TeV. Similarly, one finds in the case $N=5$ the bound
$\mu_->1.6$ TeV.
\subsection{Viable baryogenesis mechanisms}

\underline{$\bullet$ $ T_{reh}> T_{EW}$}: 
If the reheat temperature is large enough, the universe will go back
temporarily into the electroweak symmetric phase.  In this situation,
we will recover a standard cosmological evolution where eventually the
EW symmetry is broken. 

If the EW symmetry breaking proceeds by a cross-over, no departure
from equilibrium occurs and none of the common baryogenesis mechanisms
can apply here (high scale leptogenesis or EW baryogenesis).
Alternatives could be baryogenesis from out-of-equilibrium decay of
TeV scale composite states \cite{Agashe:2004bm} or low-scale
leptogenesis with TeV scale particles that produce the lepton
asymmetry by decay.  These TeV scale particles can be produced by
bubble collisions~\cite{Watkins:1991zt, Konstandin_Servant2}. In the
case of a Yukawa coupling between the fermionic species $\psi$ and the
scalar field $\phi$
\be
{\cal L} = y \, \phi \,  \bar \psi \, \psi,
\ee
a fraction $y^2$ of the energy of the scalar sector is released into
the fermions. In particular, for relativistic bubble wall velocities,
even particles with masses far beyond the electroweak scale can be
produced in bubble
collisions~\cite{Watkins:1991zt,Chung:1998ua,Chung:1998rq}. In our
scenario, almost all the energy resides in the scalar sector at the
time of bubble collisions\cite{Konstandin_Servant2}, therefore, only
very moderate couplings can produce sizable particle numbers.  It is
thus tempting to consider the possibility of TeV scale leptogenesis.
 Nevertheless, a sufficient production of right-handed neutrinos
requires Yukawa couplings that are in conflict with the generation of
light neutrino masses by the seesaw mechanism
\be
y^2 \sim \frac{m_\nu m_N}{v^2}.
\ee
In this case, the number of right-handed neutrinos can be estimated 
 ($\rho$ and $s$ denote the total energy and entropy densities)
\be
n_N \sim y^2 \frac{\rho}{m_N} 
\sim \rho  \ \frac{m_\nu}{v^2} \ \ \ \ \ \ \rightarrow  \ \ \ \ \  \frac{n_N}{s} \ll 10^{-10}.
\ee
Such value prevents a successful leptogenesis even assuming maximal CP-violating effects. 
Thus, the lepton
asymmetry has to be produced from e.g.~the decay of a new TeV species
that is not responsible for the light neutrino masses
\cite{Hambye:2001eu}. In our scenario, this is especially easy because
new TeV particles from the strongly interacting sector with ${\cal
  O}(1)$ couplings are abundantly produced non-thermally and their
subsequent decays happen far out-of-equilibrium such that washout is
avoided.

If the EW phase transition is first-order, standard EW baryogenesis is
in principle also possible and it will be interesting to investigate
further whether composite Higgs models offer the necessary conditions
(for an effective description see
\cite{Grojean:2004xa,Delaunay:2007wb}). In particular, the extended
Higgs sector in \cite{Gripaios:2009pe} seems promising. Note however
that for too strong EW phase transitions, supersonic bubbles suppress
CP violating densities in front of bubble walls, thus preventing the
mechanism of EW baryogenesis to work.

\underline{$\bullet$ $ T_{reh}< T_{EW}$}: If on the other hand the
reheat temperature is below the temperature at which the EW symmetry
is restored, a very interesting and non-trivial possibility naturally
opens up: {\it cold} electroweak baryogenesis, which is the subject of
our companion paper \cite{Konstandin_Servant2}.

We summarize the various baryogenesis possibilities in Table
\ref{tablebaryogenesis}.
\begin{table}
\begin{center}
\begin{tabular}{|c|c|c|c|c|}
\hline
& \multicolumn{2}{|c|}{ $T_{\rm reh} > T_{\rm EW}$}
& \multicolumn{2}{|c|}{ $T_{\rm reh} < T_{\rm EW}$} \\
\cline{2-5}
& EWPT is & EWPT is 
& \multirow{2}{*}{ $\left. \frac{\phi}{T} \right|_{T_{\rm reh}} > 1$ }
& \multirow{2}{*}{ $\left. \frac{\phi}{T} \right|_{T_{\rm reh}} < 1$} \\
& 1st-order & crossover &&\\ \hline
cold EW & $-$ & $-$ & $+$ & $-$ \\ 
baryogenesis & & &  &  \\ \hline
 non-local EW &  if $\phi/T |_{\rm EW} > 1$ & $-$ & $-$ & $-$ \\ 
baryogenesis & & &  &  \\ \hline
 low-scale lepto/baryogenesis & $+$ & $+$ & $-$ & $+$ \\ 
from TeV particle decays& & &  &  \\ \hline
B-conserving baryogenesis from & $+$ & $+$ & $+$ & $+$ \\ 
  asymmetric dark matter & & &  &  \\ \hline
\end{tabular}
\end{center}
\caption{
\label{tablebaryogenesis}
\small The viability of different baryogenesis scenarios depending on the
reheating temperature after the conformal phase transition and the
properties of the electroweak phase transition. $T_{EW}$ 
is the temperature at which EW symmetry gets restored. The 
last possibility is very specific as it does not require 
sphaleron processes  at any moment. It assumes that, in a 
$B$-conserving universe, the dark matter carries the anti-baryonic 
charge that is missing in the visible sector \cite{Agashe:2004bm}.
}
\end{table}

\subsection{Dark matter production during reheating 
by bubble collisions\label{sec:dark}}

The present scenario of conformal and electroweak symmetry breaking
has also important implications for the dark matter abundance. A stage
of inflation at the onset of the phase transition dilutes not only any
preexisting baryon asymmetry but also the abundance of dark matter. As
long as this era of inflation only lasts a couple of efolds, this
dilution effect is not too severe and standard scenarios of dark
matter, namely the WIMP scenario, could account for the observed dark
matter.  However, beyond around eight efolds, the preexisting dark
matter must have been overabundant before the electroweak
phase transition in order to be in accordance with today's observations,
 which is rather implausible.

There is actually no dilemma. As discussed above and in
\cite{Konstandin_Servant2}, any TeV-mass particle with significant
coupling to the radion/Higgs, hence possibly dark matter particles,
will be substantially produced at the early stages of preheating when
bubbles collide.  With a Yukawa coupling $y$, a fraction $y^2$ of the
energy in the scalar sector is transformed into dark matter particles
and already very moderate couplings of order $10^{-5}$ between the
radion/Higgs and the dark matter sector (that contains for example
stable composite states of the strongly coupled sector) will account
for the observed dark matter
abundance~\cite{Chung:1998ua,Chung:1998rq}.

\section{Experimental probes}
\label{sec:exp}
\subsection{LHC tests \label{sec:LHC}}

The underlying motivation for the framework we have been discussing
is, as well-known, that strong dynamics at the TeV scale nullifies the
hierarchy problem.  The standard realization of this scenario is
technicolor \cite{Weinberg:1975gm}, which, however, is not easy to
reconcile with EW precision measurements and flavor constraints.  In
the last years, an interesting variation interpolating between
technicolor theories and the SM Higgs model has appeared where the
Higgs emerges as a pseudo Goldstone boson from the breaking of a
global symmetry of a strongly interacting sector
\cite{Contino:2010rs}.  Generically, in this picture, we expect new
resonances at the scale $\mu_-\sim {\cal O}(1)$ TeV. However,
depending on the precise value $\mu_-$, these states may or may not be
accessible at the LHC.  A genuine strong coupling signature is the
growth with energy of the longitudinal gauge bosons scattering
amplitudes and double Higgs production. Observing these effects has
been shown to be extremely challenging and would require several
hundreds of fb$^{-1}$ of LHC data at 14 TeV \cite{Contino:2010mh}.
 
On the other hand, these models also suggest that SM fermion masses
should arise via mixing of elementary fermions with composite fermions
of the strong sector \cite{Contino:2006nn}. In this context, the top
quark is mainly a composite object while the other light SM quarks are
mainly elementary. A natural prediction is then the existence of light
($\lesssim$ 1 TeV) fermionic composite partners of the third
generation fermions, in particular the top quark
\cite{Contino:2006qr,Barbieri:2008zt,Anastasiou:2009rv}.  At the LHC, composite quarks can be
pair-produced with a large QCD cross section.  They can also be singly
produced. Prospects at the LHC for their discovery are very promising
\cite{Contino:2008hi,Mrazek:2009yu}.  There is a large number of
phenomenological studies related to this class of models, for instance
related to four-top events \cite{Pomarol:2008bh, Brooijmans:2010tn} or
signatures associated with composite leptons \cite{delAguila:2010es}.

Besides, the rate for Higgs production and decay can significantly
differ from the SM prediction.  Depending on the choice of parameters,
in particular on the ratio $v/f$, Higgs searches may either be
deteriorated or the Higgs production rate may be enhanced with respect
to the SM \cite{Espinosa:2010vn}.  Moreover, non-minimal composite
Higgs realizations typically lead to a multi-Higgs framework.  The
complexity of the scalar sector depends on the global symmetry of the
strong sector \cite{Gripaios:2009pe}.

Finally, we note that a light dilaton, as
 the pseudo-Goldstone boson of spontaneously broken scale
invariance, can fake the Higgs at the LHC.
Distinguishing it from a minimal Higgs is a subtle 
issue \cite{Goldberger:2007zk,Csaki:2007ns}. 

\subsection{A smoking gun stochastic gravity wave 
signal  \label{sec:GW}}

Gravity wave signals from the conformal phase transition have been
studied in \cite{Randall:2006py, Konstandin:2010cd}. 
The stronger is the transition, the larger is the latent heat release and therefore
the larger is the amplitude of the gravity wave signal.
The size of the signal generally scales as $(\beta/H)^{-2}$.
It should then be clear from our discussion in Section \ref{sec:PT} that an observable signal cannot be obtained without fine-tuning in the case of an ordinary polynomial phase transition, as well illustrated by Fig.~5 of \cite{Delaunay:2007wb}.
However, for a nearly conformal Goldberger-Wise  framework, several
factors favor the production of a stochastic gravity wave background
that could be observed with space-based interferometers such as LISA:
\begin{itemize}
\item Due to the large supercooling, almost all of the energy of the system
  resides in the bubble walls during the phase transition. This
  kinetic energy is essential to give rise to a sizable anisotropic stress
  that produces gravitational radiation~\cite{Randall:2006py}.
\item The phase transition proceeds rather slowly,
  c.f.~eq.~(\ref{eq:percolation}), such that typically $\beta/H
  \gtrsim O(10)$ while for generic strong phase transitions one finds
  $\beta/H \gtrsim O(100)$.
\item The peak frequency of the gravity wave spectrum is proportional  to the
  temperature of the phase transition. Since in the present scenario
  the phase transition results from the breaking of the conformal
  symmetry and hence dynamics at the TeV scale, the peak frequency is
  by about one order of magnitude larger than for a generic electroweak phase
  transition ~\cite{Huber:2007vva}. This improves the prospects for
  observation at LISA.

\end{itemize}

In conclusion, the observation of a gravity wave spectrum peaked in the millihertz range
would indicate either some sort of conformal dynamics at the TeV scale,
from a strongly interacting sector as in the
Randall-Sundrum setup (see also \cite{Jarvinen:2009mh}), or some form of low scale
inflation~\cite{GarciaBellido:2007af}.

Finally, note that while this paper has focused on the TeV scale, the
discussion can be applied to any other scale as the properties of the
phase transition do not depend on the absolute energy scale but only
on the amount of supercooling.  For instance, nearly conformal dynamics
at an intermediate scale ($\sim 10^7$ GeV), would lead to a gravity
wave spectrum peaked in the 10-100 Hz range, and could thus be probed by
LIGO\cite{Grojean:2006bp}.

\section{Conclusion\label{sec:dis}}

The framework in which EW symmetry breaking is triggered by a strongly
coupled nearly conformal sector offers an appealing dynamical solution
to the hierarchy problem.  It will take some time at the LHC to
determine whether the origin of the EW scale is due to a new strong
sector. Somehow, the cosmological consequences associated with this
scenario have not been much explored. In this paper, we have stressed
some peculiar properties of the phase transition associated with
conformal symmetry breaking and provided a study of possible
interesting cosmological features by making the least possible
reference to explicit models.

We have shown how a nearly conformal potential can lead naturally to a significant period  of supercooling. Any ordinary polynomial potential has to be fine-tuned to lead to several efolds of inflation ended by a first-order phase transition or the latter never completes. With a potential (\ref{equation:nearlyconformal}), there is no eternal inflation problem as bubbles can percolate and reheat the universe. Although the number of efolds is moderate and not sufficient to solve the horizon problem, there are still important consequences of phenomenological interest.
While we have used the Goldberger-Wise
potential for illustration, the qualitative features we have outlined
are general, in particular:
\begin{itemize}
\item A strongly first-order phase transition
\item Reheating from bubble collisions
\item A reheat temperature possibly below the sphaleron freeze-out
  temperature
\item Efficient out-of-equilibrium heavy particle (or classical field configuration) production
\item A smoking gun gravity wave stochastic background peaked in the
  millihertz range
\end{itemize}

Heavy particle production from bubble collisions was already studied
in details in the nineties. We find it somewhat appealing that the
framework we are proposing here makes this possibility quite natural
and motivates alternative cosmological scenarios from the standard
one. We refer the reader to \cite{Konstandin_Servant2} for a more
detailed discussion on reheating during bubble collisions at the TeV
scale where in particular we advocate a large production of Higgs winding configurations
and provide a description of the cold baryogenesis
mechanism in this context.

\section*{Acknowledgments}
This work is supported by the ERC starting grant Cosmo@LHC (204072)

\end{document}